\documentclass{aastex63}

\usepackage{booktabs}
\usepackage{multirow}
\usepackage{textcomp}
\usepackage{amsmath}
\usepackage[]{threeparttable}
\usepackage{lineno}
\received{Feb 10, 2021}
\revised{May 6, 2021}
\accepted{XXX}
\submitjournal{PSJ}

\shorttitle{Geophysical Exploration of Enceladus}
\graphicspath{{./}{figures/}}
\begin{document}

\title{A Recipe for Geophysical Exploration of Enceladus}

\correspondingauthor{Anton Ermakov}
\email{eai@berkeley.edu}

\author[0000-0002-7020-7061]{Anton I. Ermakov}
\affiliation{University of California, Berkeley \\
McCone Hall 307, CA 94720, USA}

\author[0000-0001-9896-4585]{Ryan S. Park}
\affiliation{Jet Propulsion Laboratory, California Institute of Technology \\
4800 Oak Grove Dr \\
Pasadena, CA 91109, USA}

\author[0000-0002-0810-1549]{Javier Roa}
\affiliation{Jet Propulsion Laboratory, California Institute of Technology \\
4800 Oak Grove Dr \\
Pasadena, CA 91109, USA}

\author[0000-0003-0400-1038]{Julie C. Castillo-Rogez}
\affiliation{Jet Propulsion Laboratory, California Institute of Technology \\
4800 Oak Grove Dr \\
Pasadena, CA 91109, USA}

\author[0000-0002-4803-5793]{James T. Keane}
\affiliation{Jet Propulsion Laboratory, California Institute of Technology \\
4800 Oak Grove Dr \\
Pasadena, CA 91109, USA}

\author{Francis Nimmo}
\affiliation{Department of Earth and Planetary Sciences, University of California, Santa-Cruz \\
Santa-Cruz, CA 95064, USA}

\author[0000-0002-1426-1186]{Edwin S. Kite}
\affiliation{University of Chicago \\
Chicago, IL 60637}

\author{Christophe Sotin}
\affiliation{Laboratoire de Planétologie et Géodynamique, \\
Université de Nantes, 44322 Nantes, France}

\author{T.~Joseph W.~Lazio}
\affiliation{Jet Propulsion Laboratory, California Institute of Technology \\
4800 Oak Grove Dr \\
Pasadena, CA 91109, USA}

\author{Gregor Steinbrügge}
\affiliation{Department of Geophysics, Stanford University \\
Stanford, CA 94305, USA}

\author[0000-0002-5126-3228]{Samuel M. Howell}
\affiliation{Jet Propulsion Laboratory, California Institute of Technology \\
4800 Oak Grove Dr \\
Pasadena, CA 91109, USA}

\author{Bruce G. Bills}
\affiliation{Jet Propulsion Laboratory, California Institute of Technology \\
4800 Oak Grove Dr \\
Pasadena, CA 91109, USA}

\author{Douglas J. Hemingway}
\affiliation{Carnegie Institution for Science \\
5241 Broad Branch Road NW \\
Washington, DC 20015, USA}

\author[0000-0002-9027-8588]{Vishnu Viswanathan}
\affiliation{NASA Goddard Space Flight Center \\
8800 Greenbelt Rd \\
Greenbelt, MD 20771, USA}
\affiliation{University of Maryland \\
Baltimore County, 1000 Hilltop Cir \\
Baltimore, MD 21250, USA}

\author{Gabriel Tobie}
\affiliation{Laboratoire de Planétologie et Géodynamique \\
CNRS/Université de Nantes, France}

\author{Valery Lainey}
\affiliation{IMCCE, Observatoire de Paris, PSL Research University, CNRS, Sorbonne Université, Univ.
Lille \\
77 Avenue Denfert-Rochereau \\
75014 Paris, France}

\nocollaboration{17}



\begin{abstract}

    Orbital geophysical investigations of Enceladus are critical to understanding its energy balance. We identified key science questions for the geophysical exploration of Enceladus, answering which would support future assessment of Enceladus' astrobiological potential. Using a Bayesian framework, we explored how science requirements map to measurement requirements. We performed mission simulations to study the sensitivity of a single spacecraft and dual spacecraft configurations to static gravity and tidal Love numbers of Enceladus. We find that mapping Enceladus' gravity field, improving the accuracy of the physical libration amplitude, and measuring Enceladus' tidal response would provide critical constraints on the internal structure, and establish a framework for assessing Enceladus' long-term habitability. This kind of investigation could be carried out as part of a life search mission at little additional resource requirements.  

\end{abstract}



\section{Introduction}

Enceladus—a cryovolcanically active and apparently habitable satellite in the Saturn system—challenges our understanding of geodynamical processes governing the evolution of ocean worlds \citep{waite2017cassini,mckay2018}. Starting in 2005, the \textit{Cassini} spacecraft revealed active eruptions in Enceladus' southern hemisphere \citep{porco2006}, confirmed the presence of a deep, global ocean \citep{iess_gravity_2014,thomas_enceladuss_2016} with complex organic molecules \citep{postberg_macromolecular_2018}, and provided direct evidence for recent hydrothermal activity \citep{hsu_ongoing_2015} that can produce redox disequilibria necessary for supporting life \citep{waite2017cassini, mckay2018}. These characteristics make Enceladus a high-priority target for astrobiology-driven exploration.

A focused geophysical investigation would underpin Enceladus' astrobiological potential. Key to understanding how Enceladus works is the spatial and temporal distribution of the energy dissipated through tidal flexing. This energy is necessary for Enceladus’ ocean to persist over geologic timescales.

The goals of this paper are:

\begin{enumerate}

    \item To provide an overview of the science questions that orbit-based geophysical data at Enceladus could answer;
    
    \item To offer recommendations for optimizing the future collection of geophysical data;
    
    \item To suggest implementation options that would address priority science questions as part of an Enceladus mission concept within NASA's medium New Frontiers class or a large Flagship class mission \citep{decadalsurvey}.
    
\end{enumerate}

\begin{figure}[t]
\includegraphics[width=18cm]{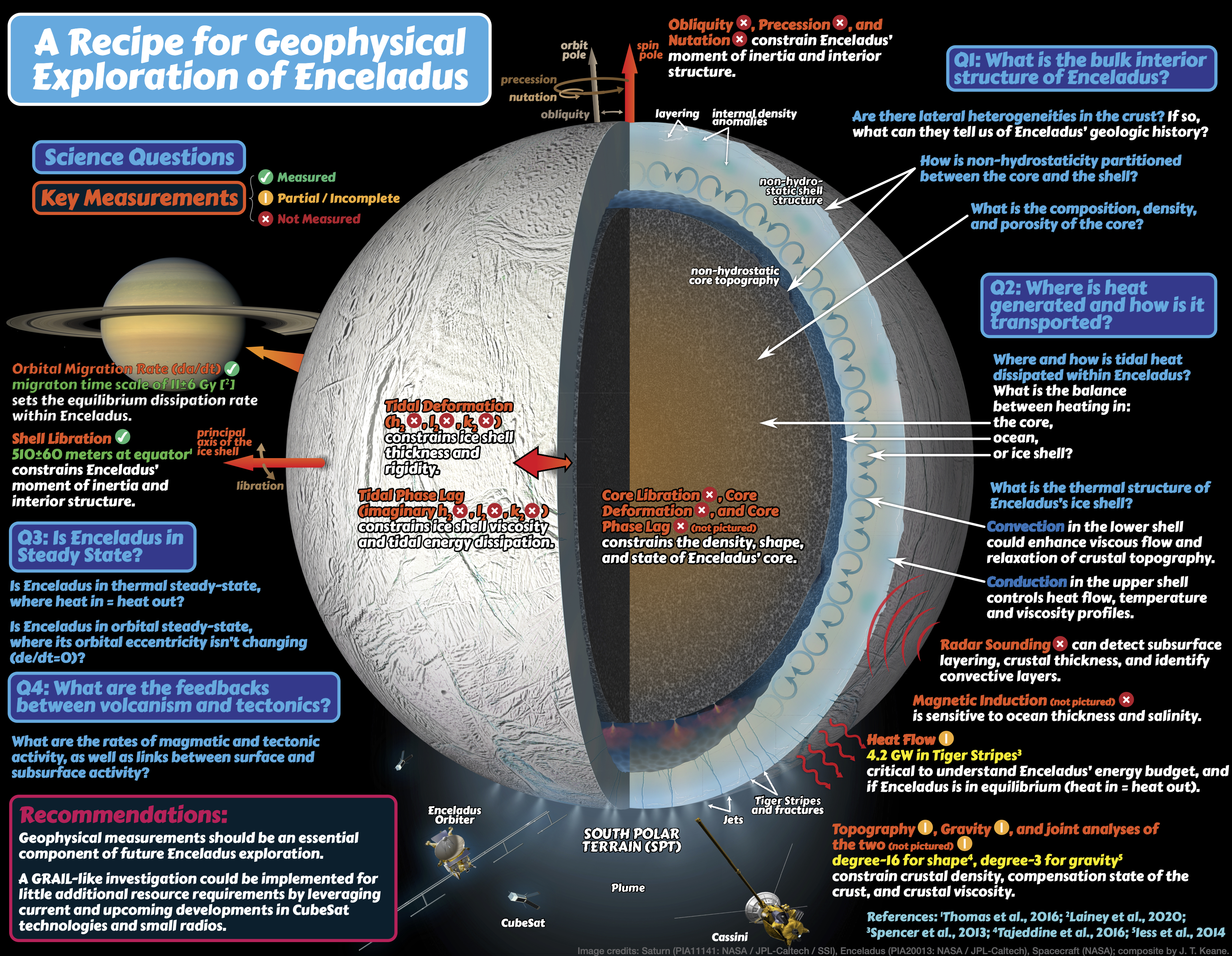}
\centering
\caption{Schematic illustration of Enceladus' internal structure along with geophysical measurements and science priority questions identified in this paper.}
\label{fig: BestFigEver}
\end{figure}

The \textit{Cassini} geophysical data, despite its limited resolution and lack of global coverage, have yielded valuable constraints on the interior state of Enceladus \citep[e.g.,][]{mckinnon2013shape,mckinnon_effect_2015,beuthe2016enceladus,hemingway_enceladuss_2019}. The gravity data along with shape models were used to constrain Enceladus' state of differentiation \citep{iess_gravity_2014,mckinnon_effect_2015} and the long-wavelength variations in ice shell thickness \citep{beuthe2016enceladus,hemingway_enceladuss_2019,cadek_long-term_2019}. The physical libration data revealed that the icy shell is decoupled from the core by a liquid layer \citep{thomas_enceladuss_2016}. The measured heat flux in the South Polar Terrain provided a lower bound on the amount of energy currently emitted from Enceladus \citep{howett_high_2011,spencer_enceladus_2013}. The coherent spatio-temporal pattern of cryovolcanic activity \citep{hedman_observed_2013,Nimmo-etal:2014} provided constraints on the rheology of and heat production within the cryovolcanically active region \citep{spitale_curtain_2015,behounkova_timing_2015,kite_sustained_2016} as well as provided hints of a longer-term variability \citep{ingersoll_time_2020}. The current state of Enceladus' geophysical data is summarized in Table \ref{tab: GeophysicalMeasurements} and illustrated in Fig. \ref{fig: BestFigEver}.

\begin{table}[ht!]
  \centering
    \begin{tabular}{p{11.835em}p{17.585em}p{23.415em}}
    \toprule
    \textbf{Quantity} & \textbf{Current Knowledge} & \textbf{Information Provided } \\
    \midrule
    \midrule
    Gravity field & Up to degree 3 field measured by Cassini \citep{iess_gravity_2014}. & Constraint on the internal structure and compensation mechanism in the ice shell. \\
    \midrule
    Shape & Estimated up to degree 16 \citep{tajeddine_true_2017}. & Constraint on the internal structure and ice shell thickness variations. \\
    \midrule
    Gravity-topography admittance$\dagger$ & Degree 3 admittance  derived \citep{iess_gravity_2014}. & Constraint on compensation state of topography, total shell thickness, elastic shell thickness and shell density and shell-ocean density contrast. \\
    \midrule
    Obliquity & Cassini-derived upper bound of 0.05$^\circ$ \citep{giese_upper_2014}. Upper theoretical bound 4$\cdot10^{-4}$ $^\circ$ \citep{baland_obliquity_2016}. & Enables determination of moment of inertia and whether Enceladus is tidally damped into a Cassini state.  \\
    \midrule
    Physical libration amplitude & Measured amplitude at the orbital frequency of 0.120$\pm$0.014$^\circ$ \citep{thomas_enceladuss_2016} or 0.155$\pm$0.014$^\circ$ \citep{nadezhdina2016libration}. & Libration amplitude is strongly sensitive to the shell thickness. Observed large amplitude requires a decoupled ice shell and hence implies a global subsurface ocean. \\
    \midrule
    Precession and nutation & Not currently measured. Precession rate estimated 2.6 rad/yr \citep{baland_obliquity_2016}. & In combination with degree 2 gravity, enables moment of inertia determination without the need for hydrostatic equilibrium assumption. \\
    \midrule
    Tidally-driven orbital migration & Measured from historical astrometric data \citep{lainey_strong_2012}. & Provides a constraint on tidal dissipation factor $Q$ of Saturn. \\
    \midrule
    Potential Love number $k_2$; radial and lateral displacement Love numbers $h_2$ and $l_2$, respectively. & Not currently measured. & Real parts are mostly sensitive to the thickness of the icy shell and its rigidity. Imaginary parts are sensitive to the viscosity profile and, thus, provide a constraint on total tidal dissipation within Enceladus. \\
    \midrule
    Radar sub-surface mapping & Not currently achieved. & Enables independent shell thickness determination. \\
    \midrule
    Magnetic induction & Not currently measured. The magnitude of the forcing field is $\approx$10~nT. & Sensitive to the sub-surface ocean conductivity and thickness. \\
    \midrule
    Thermal IR mapping & 15.8$\pm$3.1 GW from SPT \citep{howett_high_2011}; 4.2 GW from Tiger Stripes \citep{spencer_enceladus_2013}. & Constraint on the total heat flux. \\
    % add howett 2011
    \bottomrule
    \end{tabular}%
  \label{tab: GeophysicalMeasurements}
      \begin{tablenotes}
      \small
      \item  $\dagger$ Gravity-topography admittance is not an independent geophysical observable. It is derived as the ratio of the gravity amplitude to topography amplitude.
    \end{tablenotes}

      \caption{Summary of the orbit-based Enceladus geophysical data.}
\end{table}%

%   \label{tab: GeophyscialMeasurements}%
%   \caption{Summary of the geophysical data types (not in any particular order).}

Tidal dissipation within Enceladus' interior is a function of its orbital characteristics (e.g., eccentricity and proximity to Saturn) and its internal structure (e.g., thickness of the ice shell). The thermal and orbital evolutions of Enceladus are therefore coupled \citep{meyer_tidal_2007,meyer_tidal_2008,behounkova_tidally-induced_2012,neveu_evolution_2019}. That coupling has a strong effect on the long-term evolution of the satellite. The present-day internal structure determines the instantaneous spatially variable tidal dissipation rate. Depending on the efficiency of the transport of heat from the interior to the surface, Enceladus might or might not be in a thermal steady state. In that state, the heat produced within Enceladus (mostly from tidal dissipation) would equal the heat Enceladus outputs to space. In addition, tidal migration occurs, driven by dissipation within Saturn. Fast tidal migration of the Saturnian moons determined from astrometric observations indicates strong dissipation within Saturn \citep{lainey_strong_2012}. This observed fast migration implies either that the major satellites are young or that the migration rates have not been steady; the latter is predicted by the resonance locking mechanism with Saturn's normal modes \citep{fuller_resonance_2016,lainey_resonance_2020}. Thus, by studying the current internal structure and dissipation within Enceladus, we can place constraints on its tidal migration history, which is interlocked with the history of tidal heating.

We have identified the following interrelated Priority Science Questions that should be addressed by future geophysical observations of Enceladus:

\begin{enumerate}
    \item What is the internal structure of Enceladus? % Doug, Francis
    \item Where is the heat generated and how is it transported? % Julie, Francis, Gabriel, Christophe
    \item Is Enceladus currently in a steady state? % Francis, Valery 
    \item What are the feedbacks between volcanism and tectonics that regulate Enceladus’ cryovolcanism? % Edwin
\end{enumerate}

\section{Priority Science Questions }

\subsection{What is the internal structure of Enceladus?} 

Thanks to {\em Cassini}, we have a general understanding of Enceladus' internal structure \citep{hemin-etal:2018}. Gravity \citep{iess_gravity_2014} and shape data \citep{nimmo_geophysical_2011,tajeddine_true_2017} collected by \textit{Cassini}, coupled with libration data \citep{thomas_enceladuss_2016}, revealed that Enceladus has a low density, rocky core with an estimated radius of 191-198~km, overlain by a global ocean with an estimated thickness of 30-39~km and an icy shell with an estimated average thickness of 19-24~km, but thinner at the south pole \citep[4-12~km;][]{hemingway_enceladuss_2019}.

However, these inferences rely on plausible, yet untested, assumptions. In particular, the core is assumed to be in hydrostatic equilibrium, which was challenged by \cite{mckinnon2013shape} and \cite{monteux2016consequences}. The gravity-topography admittance, defined as the wavelength-dependent ratio of gravity amplitude to topography amplitude, was assumed isotropic, which would be violated if there are significant deviations from spherical symmetry. Indeed, the inferred shell thickness variations are on the order of the mean shell thickness itself \citep{hemingway_enceladuss_2019}. If these assumptions are relaxed, it becomes impossible to separate geophysical signals that originate in the icy shell (e.g., freezing/melting of the shell, convection, impact cratering, tectonics) from those originating in the core \citep[e.g., core's non-hydrostaticity and core tidal dissipation;][]{roberts2015fluffy}. Below, we summarize how a future mission could sharpen our picture of Enceladus.

\paragraph{Icy shell structure}

Currently, the mean ice shell thickness is best constrained by libration data \citep{thomas_enceladuss_2016,nadezhdina2016libration}, while shell thickness variations are deduced from gravity and topography \citep{hemin-etal:2018,hemingway_enceladuss_2019}. Improving the estimate of the libration amplitude is an easy objective for an Enceladus orbiter mission. Continuous surface observations with a stereo imager can bring down the libration amplitude uncertainty to 2~meters \citep{park_advanced_2020}---a factor of 30 improvement over the current uncertainty. The gravity and shape data determined to a higher degree would allow to extend the shell thickness inversion to smaller spatial scales, which would be especially valuable in the South Polar Terrain. The South Polar Terrain is a broad region to the south of 60\textdegree{} S. It is characterized by a 500-meter deep topographic depression \citep{nimmo_geophysical_2011,tajeddine_true_2017} and hosts a set of prominent fractures, called Tiger Stripes, that emit plumes of H$_2$O, salt, organics, and other volatiles indicating a connection to a reservoir of ocean-derived material \citep{Glein2018}. Regional gravity and topography mapping would allow to characterize the crustal structure in the South Polar Terrain. If gravity and topography data are complemented by radar sounding, an independent constraint on spatial variations in shell thickness could be derived. 

% \citep{iess_gravity_2014} published two solution for Enceladus gravity field. In both solutions, full degree~2 gravity of Enceladus has been determined. The first solution included only $J_3$ terms. The second solution estimated full degree 3 field. Most internal structure studies use the first solution. We note that the second solution has larger error bars, which is natural due to the larger number of parameters estimated. The sign of $J_3$ is uncertain in the second solution. 

Mapping gravity and topography would also be useful in more heavily-cratered terrains in Enceladus’ northern hemisphere. The relaxed state of craters in that region indicates higher heat fluxes in the past as such heating episodes would facilitate viscous flow within the shell by reducing its viscosity \citep{Bland2012}. Higher-resolution gravity data would be enable assessing the shell viscosity structure \citep{akiba2021} and would offer an independent probe of the intense past heating episode proposed by \cite{Bland2012}.

% This may be tested by measuring lateral porosity variations in Enceladus’ shell. \cite{gundlach2018sintering} concluded that Enceladus likely displays a powder-like surface consisting of unsintered micrometre-sized particles, based on the slow sintering rates observed in laboratory experiments. \cite{choukroun2020strength} also concluded that plume deposits likely remain loose and unconsolidated on Enceladus, except in the vicinity of hot spot regions. Since sintering is temperature dependent, higher temperatures from the proposed past episodes of increased heating could have led to localized sintering and variations in the shell porosity. High-resolution gravity data would be enable assessing shell porosity as high-degree gravity-topography admittance is set by the effective near-surface density \citep{besserer_grail_2014}. Correlating the inferred heat fluxes from the relaxation state of craters to the porosity variations mapped from gravity and shape data \citep[as in][]{besserer_grail_2014} would offer an independent probe of the intense past heating episode proposed by \cite{Bland2012}.

If the precession of Enceladus' pole could be measured, that would provide an independent constraint on the moment of inertia of the shell. However, the very small amplitude of the deflection (order of 1~m) would be challenging to measure, owing to Enceladus' small obliquity (see Table \ref{tab: GeophysicalMeasurements}).

\paragraph{Partitioning non-hydrostaticity between the shell and the core}

Separating the non-hydrostatic signal of the shell from that of the core could be achieved by mapping the shell thickness variations using a combination of topography, radar sounding, and gravity data. Radar sounding would enable to identify tectonics and convection zones allowing to separate the signals originating within the ice shell. A direct detection of the ice-ocean interface would allow subtracting the respective gravity signal of the ice shell and infer the properties of the core. Improving the libration amplitude measurement at multiple frequencies ranging from hours to tens of days could be used to better constrain the shell structure, specifically its moment of inertia differences (such as the difference between the shell's equatorial moments of inertia). Finally, measuring the gravitational signal of the core-shell misalignment \citep{buffett1996gravitational} could eliminate the core-shell degeneracy.  

\paragraph{Ocean thickness and composition}

Ocean thickness and density are indirectly constrained by gravity and shape data by satisfying the mass balance. The current uncertainty on the ocean thickness as constrained by gravity, shape and libration data is a factor of $\approx$2 larger than that of the shell thickness \citep{hemingway_enceladuss_2019}. The gravity-topography admittance is sensitive to the density contrast between the ocean and the shell. Thus, if the shell density is constrained from the high-degree admittance, the lower degree admittance can help constrain ocean density, which is mostly controlled by salinity. 

A measurement of the magnetic induction response convolves ocean thickness and salinity. In the Jovian system, magnetic induction has been used as a constraint on the ocean thickness \citep{kivelson2000galileo}. However, unlike Jupiter's, Saturn’s magnetic field is axisymmetric. Thus, the changing magnetic field that Enceladus experiences is primarily due to its eccentric orbit, resulting in a much smaller forcing field amplitude ($\approx$10~nT). This signal is of roughly the same magnitude as time-variable fields arising from the plumes \citep{Dough-etal:2006}. Thus, detecting an induction response and disentangling it from the plume-induced variability in the magnetic field would require long-term observations and characterization of the plume activity.

\paragraph{Rheology of the material}

The tidal response of Enceladus is described by its  Love numbers $k_{nm}$, $h_{nm}$ and $l_{nm}$ \citep[e.g.,][]{Wahr-etal:2006}, where $n$ is degree and $m$ is order. For a spherically symmetric body, the Love numbers are degenerate with respect to $m$. Thus, separating tidal signal corresponding to different orders within the same degree would indicate deviations of the body from spherical symmetry. The Love numbers are complex quantities and quantify how Enceladus' gravity field and shape respond to time-varying tidal forces. The real parts of the Love numbers depend primarily on its shell thickness and shell rigidity. The imaginary parts depend on the rheology of the material and quantify the lag in the tidal response with respect to tidal forcing, which is related to the current total dissipation within Enceladus. Independent measurement of both $h_2$ (the degree 2 radial displacement Love number, omitting the order index for simplicity) and $k_2$ (the degree 2 gravitational potential Love number) would reduce the correlation between the shell thickness and rigidity \citep[e.g.,][]{Wahr-etal:2006}. In addition, libration amplitude is sensitive to the shell rigidity in a way different from tidal deformation \citep{van2013librations}. Thus, joint measurement of the tidal response and libration would provide independent constraints on the shell thickness and its rigidity. 

One complicating factor for Enceladus is that the shell thickness varies laterally, while almost all models (including those presented in Section \ref{sec: MeasReq}) have assumed a spherically-symmetric shell when modeling the expected tidal response. \cite{A-etal:2014} and \cite{behounkova_plume_2017} are exceptions. \cite{behounkova_plume_2017} found that degree~2 Love numbers of different order could vary by a factor of two due to the nonuniform shell thickness and faults in the South Polar Terrain. 

\subsection{Where is the heat generated and how is it transported?} % Julie, Francis, Gabriel, Christophe

Tidal heating is thought to provide the main heat source, preventing quick freezing of the subsurface ocean \citep{roberts_tidal_2008}. The heat flux of Enceladus has been measured by mapping out the surface thermal emission using the Composite Infrared Spectrometer (CIRS) onboard \textit{Cassini}. The estimated value of the heat flux has varied over the years of \textit{Cassini} data analysis by a factor of several. All studies to date have focused only on the heat flux localized in the South Polar Terrain. \cite{spencer2006cassini} estimated a heat flux of 5.8$\pm$1.9~GW. A subsequent study by \cite{howett_high_2011} yielded a larger estimate of 15.8$\pm$3.1~GW using observations from a higher wavelength CIRS detector. \cite{spencer_enceladus_2013} used higher spatial resolution data and derived a total emitted power of 4.2~GW localized at the Tiger Stripes. The variability of these estimates derived from different datasets indicates the challenges of deriving the global heat flux from flyby observations of heterogeneous quality.

The conductive and advective heat fluxes are approximately equal in the South Polar Terrain \citep{kite_sustained_2016}. Advection likely dominates within the Tiger Stripes causing prominent temperature anomalies. The heat flux away from the South Polar Terrain remains unconstrained. Current estimates on global heat flux assuming a conductive shell range from 25 to 40~GW \citep{hemingway_enceladuss_2019}. Radiogenic heating within the rocky core can account for $<$0.3~GW \citep{hemingway_enceladuss_2019}, hence it is a minor contributor to the observed heat flow.

The local instantaneous tidal heat generation is determined by the product of tidal stress and strain rate. The mutual relation between these two quantities is, in turn, set by the rheology of the material that is highly sensitive to temperature. Heat is likely generated in the warmer ice at the base of the shell or in the core, especially if the core is unconsolidated (e.g., sandy/muddy). \cite{soucek_tidal_2019} used a 3D model with a variable ice shell thickness including the South Polar Terrain faults and concluded that dissipation in the solid ice cannot exceed 2.1~GW, implying that an additional heat source is needed to explain the observed heat flow. Additional dissipation could occur within the liquid slots in the ice \citep{kite_sustained_2016}. Over 10~GW can be generated by tidal friction inside the unconsolidated rocky core \citep{choblet_powering_2017}, and comparable amounts by tidal flushing of water through the porous core \citep{Liao-etal:2020}. Theoretical models predict that the ocean heat production is likely negligible \citep{Chen-etal:2014,hay_numerically_2017,rekier_internal_2019,rovira-navarro_tidally-generated_2019} because of Enceladus' low obliquity and thick ocean, although dissipation may be enhanced in narrow liquid-water conduits within the ice shell \citep{kite_sustained_2016}. However, more recently, \cite{tyler_heating_2020} indicated that there could be resonant configurations, where ocean tidal dissipation could create significant heating.

If Enceladus is in thermal equilibrium, there exists a relationship between the shell thickness and the conductive heat flux. Thus, the scientific requirement driving the uncertainty in the mean shell thickness measurement can be tied to the desired uncertainty in the heat flux. It follows from \cite{hemingway_enceladuss_2019} that in order to determine the conductive heat flux to within 10\%, the average shell thickness needs to be determined to within 2~km. In the South Polar Terrain, where the ice shell could be $<$5~km thick, radar observations may more easily reach the ice-ocean interface and return the shell thickness with an accuracy better than 100~m. This would constrain the regional dynamics and the implications for heat transport. 

% this is purely conductive shell
For a fully conductive and homogeneous shell, the thickness derived from various techniques (i.e., radar returns, tidal Love numbers, libration amplitude, or gravity-topography admittance) should be comparable. These complementary techniques offer joint advantages as the deviations from homogeneity and/or from a conductive geothermal gradient can be obtained by their cross-analysis.

Distinguishing between convection and conduction is of great interest. Conduction is likely dominant in the outer part of the shell. Convection, if it occurs within Enceladus' icy shell, would dominate heat transport in its deepest part, making the conductive region thinner and with a steeper temperature gradient, and therefore greater conductive heat loss. Efficient heat transport by convection might lead to quick ocean freezing. Large-amplitude shell thickness variations inferred from the gravity and topography data \citep{nimmo_geophysical_2011,tajeddine_true_2017,hemingway_enceladuss_2019} are at odds with a global convective layer as it would lead to fast viscous relaxation of shell thickness variations. However, a conclusive determination of whether or not convection occurs within the icy shell would require joint analysis of multiple data sets.

% obs effect of convection
A convective layer would affect the gravity-topography admittance leading to higher admittance values expected for uncompensated topography \citep{watts_isostasy_2001}. Thus, higher-resolution gravity and topography data can help identify a convective layer through their sensitivity to the viscosity profile. For a convective shell, warmer ice at shallower depth could lead to higher attenuation, especially if the salinity of the ice is high. A loss of signal resulting in no detected radar reflections at depth or weaker radar reflections from the warm and salty bottom ice would indicate a saline, convective layer. In addition, a convecting shell would lead to an apparent mismatch between the total ice shell thickness (derived from induction or gravity-topography admittance) and the thickness derived from the Love numbers, because the latter is sensitive only to the elastic part of the shell and not the weak convecting part. If the convecting part of the ice shell is less saline, assuming that parts with higher salinity would have undergone melting at some stage during convection, the ice-ocean interface could be visible as a strong reflection in the radar return. However, this direct signal could be further covered up by an accretion zone at the ice-ocean interface, which could produce ice of marine composition or result in a mushy layer with high attenuation.

%However, if convection is vigorous enough, it itself can dynamically produce topography leading to crustal thickness variations, which would also be reflected in the gravity-topography admittance. 

% In addition to viscous flow, the bottom shell topography gets eroded by melting of the thicker shell parts and refreezing onto the thinner parts. This process is caused by the pressure dependence of the freezing point of water. Water that is cooled by thermal equilibration with the ice at a higher pressure (i.e., beneath a thicker region of the shell) will, as it flows laterally to thin-shell regions, freeze at the lower pressure associated with a thinner shell since the water is now colder than the freezing point. This process will work toward evening out shell thickness variations. 

% elatic thickness
Estimating the elastic thickness would provide a critical constraint on the heat transport within the shell. Elastic thickness can be derived either from the gravity-topography admittance \citep{mcgovern_localized_2002} or by mapping flexural profiles of tectonic features that would require accurate topography knowledge \citep[e.g.,][]{giese-etal:2008}. A flexural profile amplitude of 120~m at $\approx$10-km scale is predicted in the vicinity of Tiger Stripes \citep{hemingway_cascading_2020}. Deriving spatial variations of the elastic thickness and cross-correlating them with the heat flow mapped in the thermal IR emission could help validate the dissipation pattern within the shell. Measuring flexural signals would require regional topography knowledge to at least 10-m vertical accuracy, which would require a dedicated stereo mapping campaign. Stereo-derived digital terrain model can be further improved and geodetically referenced to the center-of-mass frame by laser or radar altimeter data.

\subsection{Is Enceladus currently in a steady state?} % Valery, Francis
% Q/2pi is the ratio between the total energy of an oscillation and the energy dissipated per cycle
The long-term orbital evolution of Enceladus depends on both the dissipation within Saturn and Enceladus. The dissipation is described by the so-called quality factor $Q$, which is proportional to the ratio of energy stored in tidal motion to the energy dissipated over one tidal cycle. $Q$ depends on the frequency of tidal forcing \citep[e.g.,][]{wu2005origin} and thus, can change over the tidal migration history. In addition, $Q$ may evolve as the temperature of the body evolves due to secular cooling. Recent astrometric efforts indicate that Saturn is more dissipative than previously thought. \cite{lainey_strong_2012} found Saturn's $Q$ as small as $\approx$2,000 at Enceladus' tidal frequency, permitting equilibrium tidal dissipation within Enceladus to be as much as $\approx$25~GW \citep{meyer_tidal_2007}. 

The feedback between orbital and thermal evolution leads to two distinct kinds of steady state. First, in the thermal steady state, the present-day heat production is equal to the present-day heat loss. Second, in an orbital steady state, damping of eccentricity, $e$, due to dissipation within Enceladus is balanced by pumping of its eccentricity by the 2:1 resonance between Dione and Enceladus (i.e., $de/dt=0$; \citet{meyer_tidal_2007}). If not in a steady state, Enceladus could exhibit a periodic behavior \citep{OjakaSteve:1986}, in which energy is stored at one time and released at a later time (perhaps resulting in cyclical variations of ocean thickness), or it might be in a net freezing or melting state. Indeed, topographic evidence for viscous relaxation of impact craters in ancient terrains \citep{Bland2012} and mapping of fault patterns away from the South Polar Terrain \citep{Patterson2018} both show that the location and/or intensity of high crustal heat flow and associated tectonic activity has changed over geologic time. Geophysical investigation of these tectonic anomalies may provide clues as to non-steady-state behavior in Enceladus' past: for example, the processes that drove the onset of hyperactive resurfacing near the South Pole, but not the North Pole \citep[e.g.,][]{KangFlierl2020}.

Both kinds of steady states critically depend on the tidal phase lag, which is characterized by the imaginary part of the potential Love number,~$\mathrm{Im}(k_2)$. Larger values of $\mathrm{Im}(k_2)$ correspond to larger internal tidal dissipation (see Eq. \ref{eq: TidalDissipation}) and faster tidal damping of eccentricity. $\mathrm{Im}(k_2)$ can be derived in two ways. First, a tidal phase lag would affect the orbit of the spacecraft: the measured response would lag that expected from the perturbing potential. Thus, it can be derived by radio-tracking of the spacecraft in the same way the gravity field is derived. Second, the tidal phase lag affects the orbit of Enceladus, causing the damping of its eccentricity. Thus, it could be derived by determining an accurate ephemeris model using ground- or spacecraft-based data (e.g., radio ranging or astrometry). Pursuing both ways of deriving the tidal phase lag would provide a robustness check. In addition, the change of eccentricity should be accompanied by the corresponding evolution of the Enceladus-Dione resonance libration angle, the determination of which would require precise ephemerides of both Enceladus and Dione. 

In conclusion, long-baseline continuous ground-based astrometry at the kilometer level and future radio ranging and high-accuracy astrometry in the Saturnian system (20~years after \textit{Cassini}) would be needed to reveal whether Enceladus is in the orbital steady state. Thermal IR mapping of Enceladus, including the conductive heat flux away from the Tiger Stripes, is needed to assess if Enceladus is in the thermal steady state. 

\subsection{What are the feedbacks between volcanism and tectonics that regulate Enceladus’ cryovolcanism?} % Edwin

Enceladus cryovolcanism leads to significant mass loss from the moon \citep[150--350 kg/s or 4-10\% of Enceladus' mass per Gy,][]{hansen2006enceladus} and also offers the opportunity to sample material from Enceladus' ocean. The rate of surface-interior exchange is important for sustaining habitability \citep[e.g.,][]{Soderlund2020}. However, the causes of the tectonic features of the volcanically active South Polar Terrain are not well understood \citep[e.g.,][]{yin_gravitational_2015,hemingway_cascading_2020}, nor are the resurfacing mechanisms or rates of resurfacing and surface-interior exchange well constrained \citep{Spencer2018, Bland2015}. As a result, we do not know whether the present-day rate of cryovolcanic activity is representative of the history of the South Polar Terrain \citep{ONeillNimmo2010}.

Enceladus operates in a regime intriguingly different from the other worlds known to have active volcanism such as the Earth and Io. Earth's oceanic lithosphere is dominantly cooled by conduction (ratio of advected to conducted heat $\approx$0.1). On Io, volcanism drives the cooling and, thus, the tectonics (ratio of advected to conducted heat $\approx$10). By contrast, on Enceladus, the ratio of volcanic heat to conducted heat appears to be $\approx$1, implying that the South Polar Terrain evolution cannot be understood without considering both processes. The strong potential coupling between volcanism and tectonics highlights the importance of a high-resolution gravity and topography mapping within the South Polar Terrain as well as higher spatial resolution mapping of heat flow. Estimates of the elastic thickness from the flexural profiles collected in the vicinity of the cracks might be significantly different from the estimates using localized gravity-topography admittance since a significant advection of heat occurs in the fractures. Better constraints on the spatio-temporal pattern of eruptions \citep{Nimmo-etal:2014,spitale_curtain_2015} might constrain the heat production within the Tiger Stripes and the plumbing system, as well as tectonic mechanisms operating within the South Polar Terrain \citep{kite_sustained_2016}. High-resolution mapping of how surface thermal emission falls off with distance from the Tiger Stripes would provide additional constraints on the thermal structure of the plumbing system \citep[e.g.,][]{AbramSpence:2009}.

Radar sounding would provide a direct way to investigate tectonic activity. Faulting processes can leave dielectric signatures leading to strong reflectors in the radar return. Furthermore, compositional variations associated with tectonic activity, injection of water, or even variations in crystalline fabric can be detected in radar sounding data. Vertical fractures could be identified by point scattering usually leaving characteristic hyperbolas in the (non-focused) radar return. Due to the complex surface topography of Enceladus, the radar sounding data has to be complemented by high-resolution topography data to mitigate surface clutter.

Higher-resolution gravity, shape and heat flux as well as radar sounding and data within the South Polar Terrain would open the prospect of studying a new kind of tectonics as rich and unusual by terrestrial standards as plate tectonics when first understood in the 1960s and 1970s. High-resolution gravity (50-km resolution, or $\approx$degree~30, to fully resolve the South Polar Terrain) in combination with stereo digital elevation models could be particularly valuable. Crucial constraints on the rates of both magmatic and tectonic activity, as well as links between surface and subsurface activity, might be provided by Interferometric Synthetic Aperture Radar (InSAR) measurements---analogous to what has been done successfully for Earth such as mapping the ice flow velocities over Antarctica \citep{Rignot2011} or mapping active volcanism-induced deformations \citep[see][and references therein]{Segall2010}.

%such as gravity signatures of co-seismic and post-seismic deformation due to the rupture of 2004 Sumatra-Andaman earthquake \citep{chen2007grace}, or 
% Comment from Edwin: I think the broader point is we don't know what is going on and high-resolution gravity could help build intuition (just as it did for figuring out plate tectonics on Earth)
% This comments needs to be quantified

\section{Development of measurement requirements}\label{sec: MeasReq} % Anton 

Geophysical measurement requirements can be derived in multiple ways as several combinations of measurements can yield identical accuracy for a recovered parameter. For example, gravity and radar measurements can both yield shell thickness estimates. In addition, the measurement requirements might be dependent on the (yet unknown) value of the recovered parameter. For example, the libration amplitude is an inverse function of shell thickness.

In order to develop traceable measurement requirements, we used the Markov chain Monte-Carlo (MCMC) approach to develop a framework for connecting the science requirements to the measurement requirements \citep{matsuyama_grail_2016}. We use central values for the internal structure parameters (such as layer thicknesses and densities) from \cite{hemingway_enceladuss_2019} as the truth values. In addition, we assumed a prior probability distribution for all model parameters, including the currently unconstrained (due to a lack of Love number measurements) viscoelastic moduli. The form of the prior probability distribution is summarized in Table \ref{tab: MCMC_prior}. 

\begin{table}[ht]
  \centering
    \begin{tabular}{lccl}
    \toprule
    Parameter & \multicolumn{1}{l}{Minimum value } & \multicolumn{1}{l}{Maximum value} & Distribution type \\
    \midrule
    \midrule
    Shell shear modulus [GPa] & 1     & 10    & log-uniform \\
    \midrule
    Shell Poisson's ratio & 0.25  & 0.45  & uniform \\
    \midrule
    Shell viscosity [Pa s] & 10$^{14}$ & 10$^{19}$ & log-uniform \\
    \midrule
    Shell density [kg/m$^3$] & 600   & 1100  & uniform \\
    \midrule
    Shell thickness [km] & 2     & 40    & uniform \\
    \midrule
    Ocean compressibility [GPa] & 2     & 3     & log-uniform \\
    \midrule
    Ocean density [kg/m$^3$] & 920   & 1300  & uniform \\
    \midrule
    Ocean thickness [km] & 2     & 80    & uniform \\
    \midrule
    Core shear modulus [GPa] & 10    & 100   & log-uniform \\
    \midrule
    Core Poisson's ratio & 0.25  & 0.45  & uniform \\
    \midrule
    Core viscosity [Pa s] & 10$^{20}$ & 10$^{22}$ & log-uniform \\
    \midrule
    Core density  [kg/m$^3$] & 2000  & 3000  & uniform \\
    \midrule
    Core radius [km] & 2     & 252   & uniform \\
    \bottomrule
    \end{tabular}%
   \label{tab: MCMC_prior}
  \caption{Prior probability distribution of Enceladus internal structure parameters. In addition to these bounds, the densities were constrained to increase monotonically with depth. The thickness of the three layers were constrained to add up to the outer radius of Enceladus.}
  \end{table}%

%   \label{tab: MCMC_prior}
%   \caption{Prior probability distribution of Enceladus internal structure parameters. In addition to these bounds, the densities were constrained to increase monotonically with depth.}

% 

To generate synthetic observations, we compute a multi-layer hydrostatic equilibrium model using \cite{tricarico2014multi}. Viscoelastic moduli and densities are assumed constant within each layer. This gives us the hydrostatic shape and gravity spherical harmonic coefficients. Degree~2 complex Love numbers are computed in a way similar to \cite{kamata2015tidal}. Enceladus is assumed to have a small non-hydrostatic topography, which is compensated with an Airy isostasy mechanism \citep{watts_isostasy_2001}. This yields modeled gravity-topography admittance. The correlation between gravity and topography is assumed to be unity. The libration amplitude at the orbital frequency is computed for rigid shells \citep{van2008librations}, which is a simplification of our model. As the libration measurement accuracy is improved, it becomes sensitive to the rigidity of the shell. \cite{cadek2016enceladus} estimated that varying the shell rigidity between $10^9$ and $5\cdot10^9$ Pa leads to a libration amplitude difference of 50 meters, which is just below current observational uncertainty. The variation of libration  amplitude due to the shell viscosity profile is yet smaller: $<$ 20 meters for bottom shell viscosity ranging from $10^{12}$ to $10^{15}$ Pa s \citep{vanHoolst2016diurnal}. More details on the MCMC geophysical inversion are given in Appendix \ref{sec: MCMC}.

In assuming the measurement uncertainties, we impose that future measurements of physical libration and gravity coefficients will not be worse than the current ones \citep{iess_gravity_2014,thomas_enceladuss_2016}. We explore the combined sensitivity of physical libration amplitude at the orbital period; tidal Love numbers $k_{2}$, $h_{2}$, $l_{2}$; and the gravity-topography admittance spectrum $Z_l$, where $l$ is the spherical harmonic degree, to the internal structure parameters. For each set of measurement errors, we derive the posterior distribution of internal structure model parameters. In order to visualize the multidimensional posterior distribution, we marginalize it over the parameters of interest (e.g., shell thickness or shell density).

\begin{figure}[ht]
\includegraphics[width=\columnwidth]{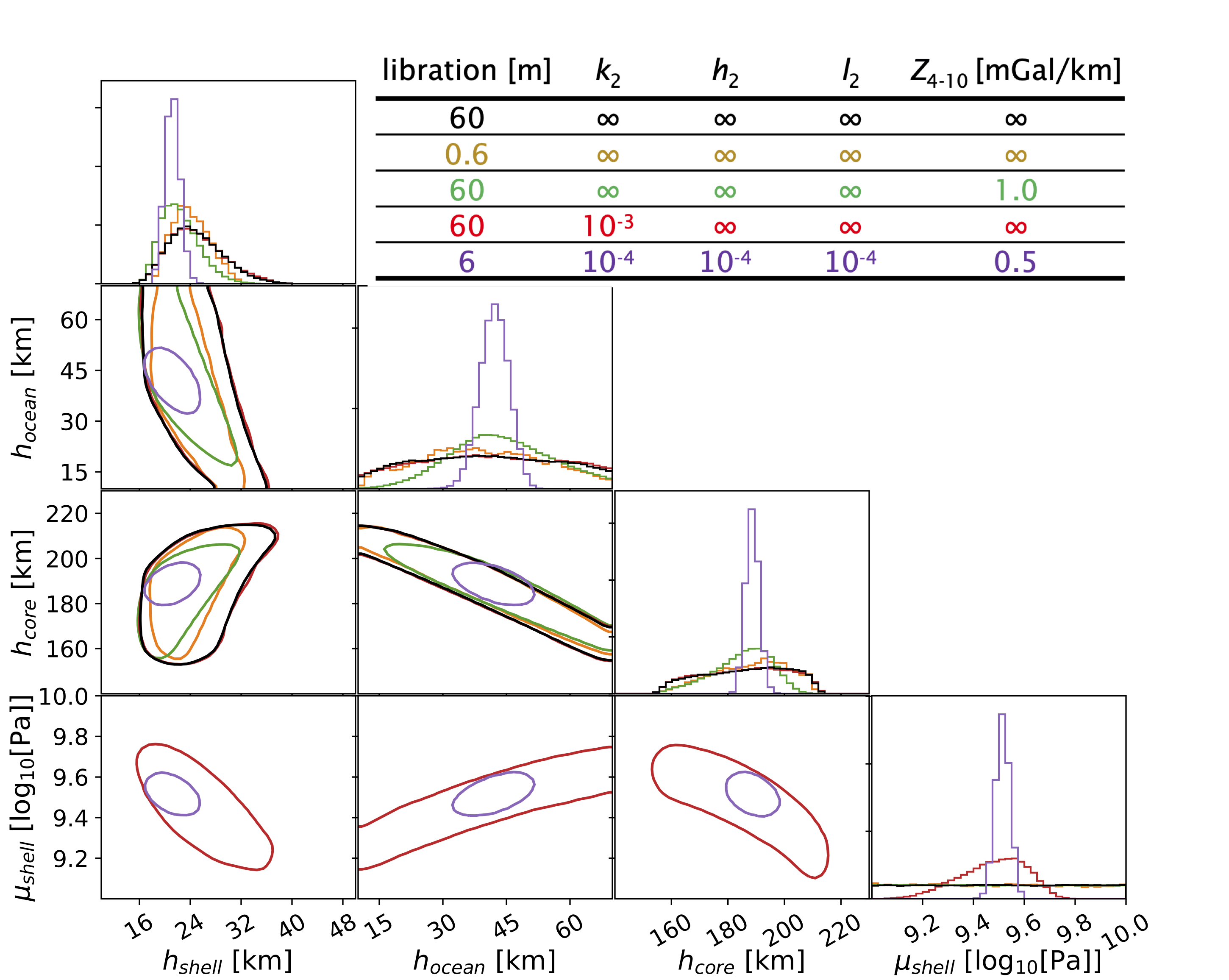}
\centering
\caption{Corner plot visualizing the posterior distribution of internals structure model parameters. Histograms for layer thicknesses ($h_i$) and shear modulus of the icy shell ($\mu_{shell}$) are shown. Note that $\mu_{shell}$ is shown on a logarithmic scale. $\mu_{shell}$ is constrained only when Love numbers are measured. The contours show the 95\% percentile. The black contour corresponds to the current state of knowledge, while the colored contours show how the posterior distribution changes depending on the additional measurements included in the inversion. $Z_l$ stands for the gravity topography admittance at degree~$l$. The table at the top shows the measurement uncertainties for the corresponding contours. The infinity sign in the table indicates that the quantity was not included in the inversion.}
\label{fig: MCMC_figure}
\end{figure}

We have explored several combinations of measurement uncertainties and mapped them with MCMC into the internal structure parameter posterior distributions. Fig. \ref{fig: MCMC_figure} shows the posterior distributions for selected internal structure parameters in form of a corner plot \citep{foreman2016corner}. The top boxes show the 1D histograms of parameters of interest. In addition, we show 95\% confidence regions for selected parameter pairs. The black contours in Fig. \ref{fig: MCMC_figure} represent the current state of knowledge. The other (colored) contours show how the confidence regions shrink as more data is included in the MCMC inversion.

% Anton to present several representative corner plots

We find that improving the accuracy of admittance (mostly) and libration (less so) can decrease the shell thickness uncertainty. Improving the accuracy of the gravity-topography admittance to the level of 1~mGal/km up to degree~10 and libration amplitude to an accuracy of 6~meters can decrease that uncertainty down to 2~km, which corresponds to a 10\% uncertainty in the conductive heat flux. The Love numbers provide sensitivity to viscoelastic moduli. If $k_{2}$ is measured to $10^{-2}$ or better, it provides a constraint on shell rigidity and improves the accuracy of the shell thickness determination. If $k_{2}$ is measured to $10^{-3}$, it becomes sensitive to the shell viscosity. 

\begin{figure}[ht]
\includegraphics[width=\columnwidth]{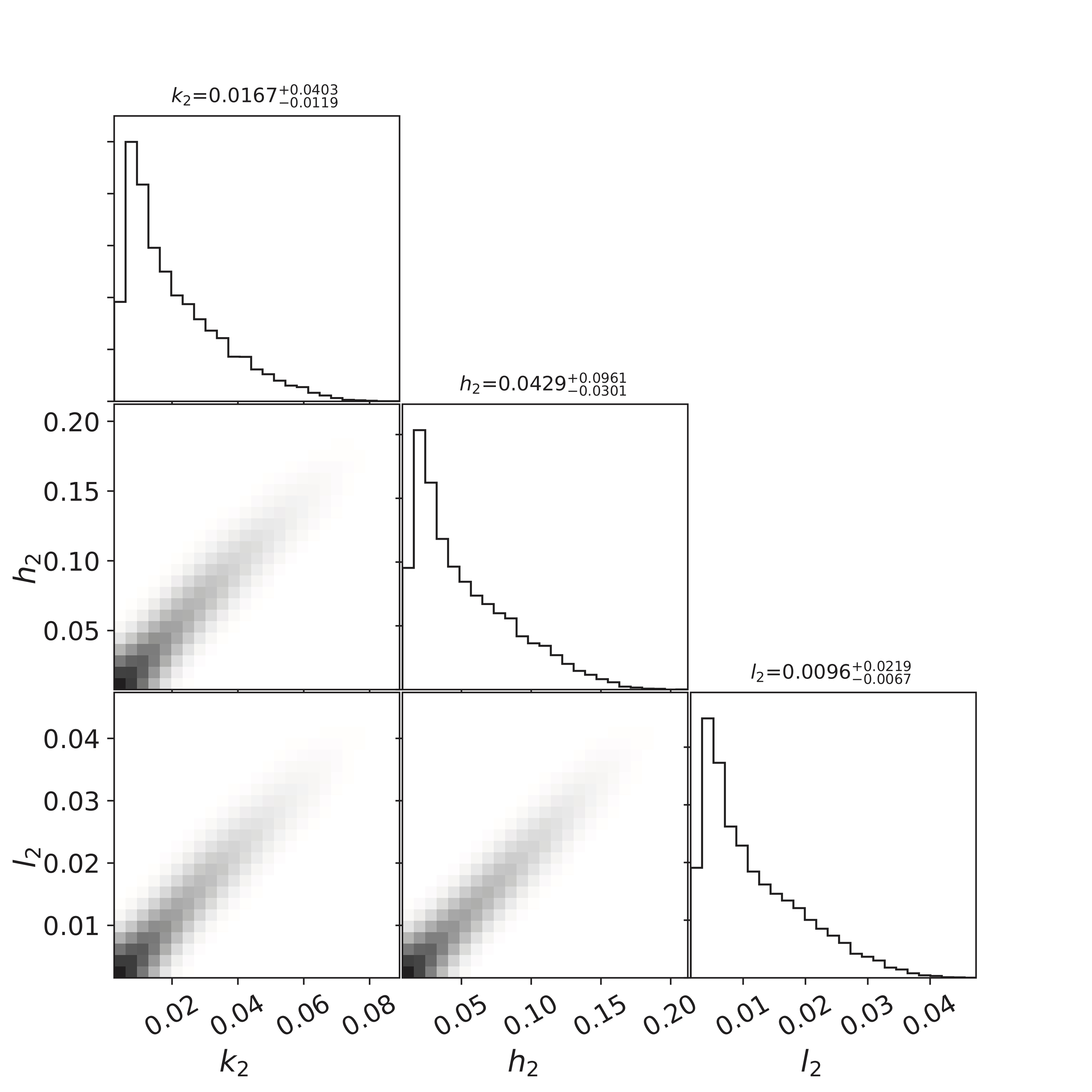}
\centering
\caption{Corner plot visualizing the posterior distribution of Enceladus' real parts of Love numbers as constrained by the gravity \citep{iess_gravity_2014}, shape \citep{tajeddine_true_2017} and libration data \citep{thomas_enceladuss_2016}. The darker colors in the 2D histograms indicate higher probability values. The 2.5\%--97.5\% percentile ranges are given at the top.}
\label{fig: LoveNumberPosterior}
\end{figure}

\subsection{Estimating the amplitude of tidal deformations}

Having estimated  Enceladus' Love numbers using an MCMC inversion, we can statistically assess the magnitude of surface tidal deformation. The posterior distribution for the Love numbers is shown in Fig. \ref{fig: LoveNumberPosterior}. The derived confidence intervals for the Love numbers allow an estimation of the tidal displacement ranges and surface gravity changes to be measured by future spacecraft missions. The tidal surface displacement is proportional to the displacement Love numbers $h_2$ and $l_2$. The maximum range of tidal displacement from degree~2 tides is given by \cite{park_advanced_2020} as:

\begin{eqnarray}
\max(\Delta R) = 12\sqrt{\frac{2}{7}} \frac{h_{2} R^2 \omega^2 e}{g}\\
\max(\Delta N) = 12\frac{l_{2} R^2 \omega^2 e}{g}\\
\max(\Delta E) = 9 \frac{l_{2} R^2 \omega^2 e}{g}
\label{eq: TidalDisplacementRange} 
\end{eqnarray}

\noindent for the radial, northerly and easterly displacement, respectively. In addition, the maximum surface gravity variation range is given by:

\begin{equation}
   \max(\Delta g) = 24\sqrt{\frac{2}{7}}(1 - \frac{3}{2}k_2 + h_2) R \omega^2 e
\label{eq: SurfaceGravityRange} 
\end{equation}

% We used Markov chain with current measurement uncertainties and explored a range of rheologic parameters to derive a posterior distribution of the Love numbers, which is shown in Fig. \ref{fig: LoveNumberPosterior}

\begin{figure}[ht]
\includegraphics[width=\columnwidth]{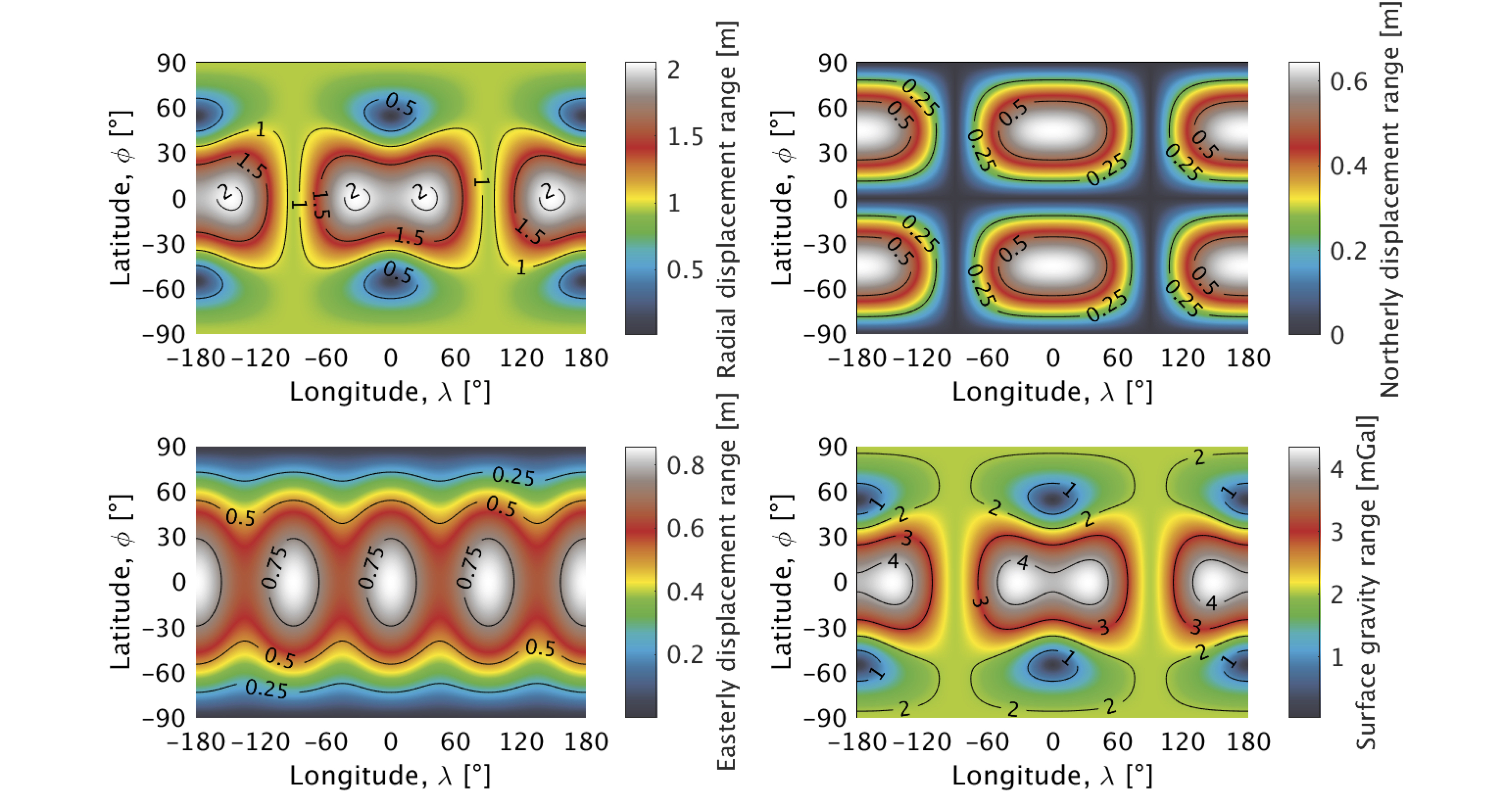}
\centering
\caption{Radial, northerly and easterly tidal displacement ranges as well as surface gravity range of Enceladus given the median values of the Love numbers as constrained by the gravity \citep{iess_gravity_2014}, shape \citep{tajeddine_true_2017} and libration data \citep{thomas_enceladuss_2016}.}
\label{fig: TidalDispalcementRanges}
\end{figure}

\noindent Taking the median values for the Love numbers from Fig. \ref{fig: LoveNumberPosterior} ($k_{2}$ = 0.0167, $h_{2}$ = 0.0429, $l_{2}$ = 0.0096), we derive the maximum range of tidal displacement of 2.0, 0.9 and 0.6 meters in the radial, northerly and easterly directions. The map of tidal displacement ranges for these median values of Love numbers is shown in Fig.~\ref{fig: TidalDispalcementRanges}. In addition, Fig.~\ref{fig: TidalDispalcementRanges} shows the range of surface gravity changes, which has the same pattern as the vertical tidal displacement range. It can be seen that in our spherically symmetric model tidal displacements are maximized at the equator. 

The maximum radial tidal deformation range is 0.5--6.5 meters (2.5--95\% confidence interval) in the equatorial region but can be amplified in the vicinity of the Tiger Stripes \citep{behounkova_plume_2017,Marusiak2021OWGeophys}. We note that tidal displacements are expected to be much smaller compared to the libration amplitude at the orbital frequency \citep[$\approx$530 meters at the equator,][]{thomas_enceladuss_2016}. Thus, an accurate libration model would be required to tease out tidal deformations. The combined measurement of the gravity-topography admittance, libration amplitude and tidal deformation can effectively reduce the shell thickness uncertainty. Finally, measuring obliquity and precession would require a sub-meter accuracy on the Enceladus orientation.

Given the observed heat flux \citep{howett_high_2011}, we can derive an estimate of the imaginary part of the potential Love number using the tidal dissipation formula \citep{peale1979melting}:

\begin{equation}
\dot{E} = \mathrm{Im}(k_2)\frac{21}{2}\frac{GM_{Saturn}^2 R_{Enceladus}^5 n e^2}{a^6}.
\label{eq: TidalDissipation} 
\end{equation}

\begin{figure}[ht]
\includegraphics[width=\columnwidth]{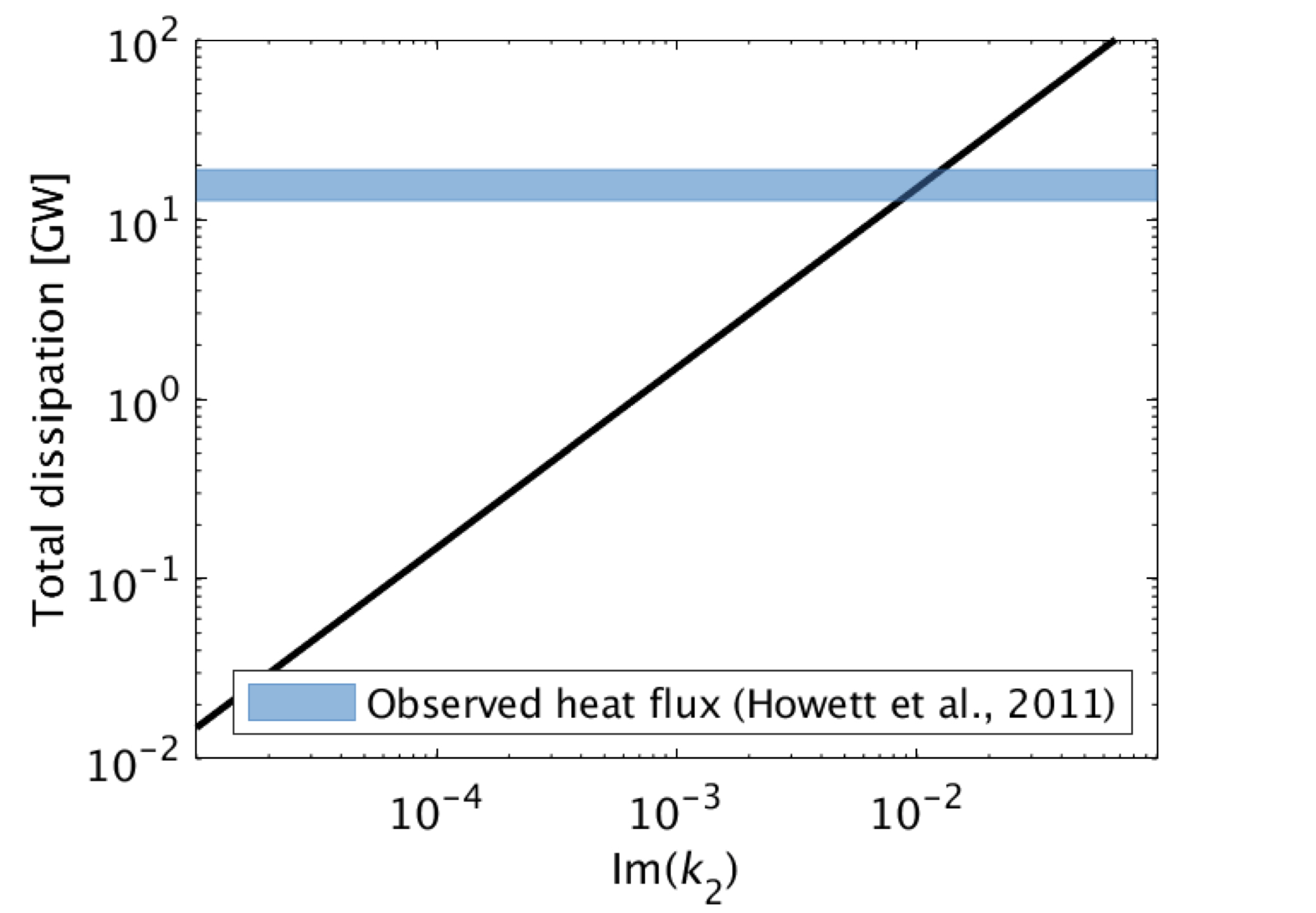}
\centering
\caption{Tidal dissipation of Enceladus as a function of the imaginary part of the Love number.}
\label{fig: TidalDissipation}
\end{figure}

\noindent Thus, the measurement requirement on the $\mathrm{Im}(k_2)$ should be tied to the requirement of on the heat flux measurement. Fig. \ref{fig: TidalDissipation} shows the total tidal dissipation as a function of $\mathrm{Im}(k_2)$, from which follows that measuring $\mathrm{Im}(k_2)$ to $10^{-3}-10^{-2}$ would be required for detecting the tidal lag and constrain the total dissipation, assuming Enceladus is currently in thermal equilibrium. Finally, the recovery of $h_2$ and $l_2$ with the same accuracy as for $k_2$ can help mitigate the ambiguity of the ice rheology that arises when measuring $k_2$ only \citep{Wahr-etal:2006}.

\section{Mission design considerations}
\subsection{Gravity and tides sensitivity} % Anton, Bruce, Javier, Ryan.
% this subsection will describe how the tidal Love number recovery accuracy depends on mission configuration. 

We have studied the sensitivity of single spacecraft with Earth-based Doppler tracking and dual spacecraft with inter-spacecraft tracking (i.e., GRACE/GRAIL-like configuration, \cite{tapley2004gravity,zuber2013gravity}) to map the static and temporally-variable gravity field of Enceladus. Polar orbits with global coverage are preferred for global static gravity mapping. However, such orbits at Enceladus are unstable and other special types of high-inclination orbits need to be considered \citep{russell_design_2009,massarweh2020restricted}. High-inclination orbits are also of interest for gravity mapping of the South Polar Terrain of Enceladus that lies below 60\textdegree{} S.

Tidal deformation causes the gravity field of Enceladus to vary in time, which affects the motion of spacecraft around it. When operating a formation of multiple spacecraft around Enceladus, the tidal investigation would rely both on the effect of tides on each spacecraft but also on the differential effect of the perturbation, which affects the spacecraft relative motion. The effect of tides on the relative motion depends on the relative position of the spacecraft. Assuming that a formation of two spacecraft co-orbits Enceladus at the same mean altitude along circular equatorial orbits, the design space reduces to choosing their angular separation and orbit altitude. The effect of tides on the motion of a spacecraft can be split into two components:

\begin{enumerate}
    \item Direct (short-term): tides cause short-period variations in the orbital elements of a single spacecraft or periodic changes in the range-rate between two spacecraft over one orbital revolution.
    
    \item Indirect (long-term): tides can cause long-term variations in the orbital elements or long-term changes in the range-rate between two spacecraft. The indirect effect may be amplified by resonances between spacecraft's mean-motion and tidal harmonics.
    
\end{enumerate}

% Tidal deformation occurs at specific frequencies. We found that the effect of tides on a spacecraft orbit is amplified when the mean motion of the spacecraft is in 3:1 and 1:1 resonance with the orbital frequency of Enceladus. These resonances can potentially amplify the sensitivity to tides. At Enceladus, the orbit altitudes leading to the resonant frequencies are unstable due to Saturn's perturbation. However, this finding could be useful for other tidally-locked bodies. 

The amplitude of the tidal deformation and tidal gravity perturbation is at the maximum at the equator (Fig.~\ref{fig: TidalDispalcementRanges}). For this reason, equatorial orbits are in general better for optically observing tidal deformation and detecting tidally induced gravity variations as they would maximize the observable signal. In addition, displacement due to forced libration is also maximized at the equator.

If the body is spherical, the Love numbers are degenerate with respect to spherical harmonic order $m$. Thus, for such a body: $k_{20}=k_{21}=k_{22}$. Deviations from spherical symmetry would break this degeneracy. \cite{behounkova_plume_2017} estimated that Enceladus' degree~2 potential Love numbers could be different by a factor of two. In order to separate the effects of the degree~2 Love numbers, a non-equatorial orbit would be needed. Thus, the choice of an optimal orbit should depend on the expected internal structure and should be driven by a hypothesis to maximize the sensitivity to the phenomenon in question. 

\subsection{Orbital stability} % Ryan, Javier
% this subsection will describe the stability of spacecraft orbits around Enceladus, that would inform future mission design

% first talk about a single spacecraft
The stability of orbits around Enceladus is driven by the tidal force due to Saturn. Orbital stability in this context refers to a spacecraft remaining in orbit without escaping or impacting Enceladus. In practice, a certain orbit may be considered stable as long as its lifetime, although finite, is long enough to complete the scientific exploration phase according to the mission requirements.

Tidal perturbations due to Enceladus' tidal deformation are small compared to the perturbation from Saturn---even compared to the third-order terms of the static gravity field---and do not pose a risk to orbital stability. For this reason, it is relevant to find the orbit configurations that maximize the tidal signature when designing a mission to recover the Love numbers.

Low-altitude, near-circular orbits are a versatile option suitable for achieving different scientific goals, including gravity recovery and surface mapping. However, such orbits are only stable at relatively low inclinations and low altitudes over Enceladus. The Lidov-Kozai (LK) mechanism \citep{1962P&SS....9..719L,1962AJ.....67..591K} produces a long-period exchange between inclination and eccentricity, effectively increasing the eccentricity of the orbit while reducing its inclination and vise versa. Assuming that the radius of the spacecraft orbit is small compared to the radius of the orbit of Enceladus, a first-order expansion of the three-body potential leads to the quadrupole approximation, which predicts the maximum eccentricity during an LK cycle to be:

\begin{equation}\label{Eq:eccentricity}
    e_\text{max} = \sqrt{1 - \frac{5}{3}\cos^2i},
\end{equation}

\noindent where $i$ is the orbital inclination. This approximation is only valid for $\cos^2i<3/5$, or for inclinations between the Kozai angles $i_\text{min}=39.2$\textdegree{} and $i_\text{max}=140.8$\textdegree. The eccentricity of initially circular orbits with inclinations between these critical values would grow due to the LK mechanism, effectively lowering its periapsis eventually leading to an impact. Given the semimajor axis of the orbit, Eq.~\eqref{Eq:eccentricity} can be used to estimate the inclination at which the periapsis reaches the surface of Enceladus. However, \citet{lara2010mission} proved that the quadrupole expansion is not sufficient to accurately model the dynamics at Enceladus, and found that a higher-order expansion predicts stable near-circular orbits at inclinations as high as 50\textdegree.

The maximum altitude that low-inclination near-circular orbits can reach while remaining stable is driven by the Coriolis asymmetry \citep{1979AJ.....84..960I}. When modeling orbital motion in the Saturn-Enceladus synodic frame, the Coriolis effect opposes the gravitational attraction from Enceladus for prograde motion whereas it supplements Enceladus' gravity for retrograde motion. As a result, the maximum stable orbital radius of prograde orbits is only half of that for retrograde orbits. Stable prograde circular equatorial orbits are stable below altitudes of 200~km, whereas their retrograde counterparts are stable up 700~km above the surface.

An alternative design approach is to seek periodic orbits in the Saturn-Enceladus circular restricted three-body problem (CR3BP), which may present larger eccentricities and also reach higher inclinations \citep{russell_design_2009}. The different families of periodic orbits in the CR3BP provide mission planners with a wide variety of options for pursuing specific objectives. For example, the low-altitude near-circular equatorial orbits already discussed are members of the family of distant prograde orbits (DPOs) and distant retrograde orbits (DROs), respectively. The continuation of the family of periodic DPOs yields orbits with different geometries but these are also unstable. \citet{russell_design_2009} noted that northern halo orbits, which are eccentric and highly inclined, provide unique opportunities for exploring the South Polar Terrain and observing the Tiger Stripes, a configuration similar to Dawn's Second Extended Mission \citep{park_evidence_2020}. A subset of these halo orbits is stable with periapsis altitudes ranging from 30~km all the way down to the surface. Although stable in the simplified CR3BP model, the actual orbit lifetime of this set of halo orbits is less than one week. \citet{davis2018trajectory} presented additional periodic orbits with interesting opportunities to observe the South Polar Terrain. Although such highly inclined orbits tend to be unstable in the full-ephemeris model, \citet{davis2018trajectory} demonstrated that halo orbits flying over the South Polar Terrain at altitudes lower than 200~km can be controlled with an estimated $\Delta V$ cost of 20~m/s per month. Nuclear electric propulsion can significantly increase the maneuvering capabilities for orbit transfer and control \citep{casani2020enabling}, potentially enabling a wider range of science orbits.

%Equatorial orbits are most attractive for tidal investigation because the maximum radial and longitudinal variations of Enceladus' gravity occur in the equatorial region. However, equatorial orbits are not well suited for investigations of the static gravity field and surface imaging, where global coverage with high-inclination orbits are preferred.

A mission consisting of more than one spacecraft requires not only the orbit of each spacecraft to be stable, but also ensuring that the relative formation is preserved. The potential of formation flying concepts for deep-space exploration has been discussed in the past, including design and control considerations \citep{gurfil2003adaptive,howell2005natural}. Examples of proposed applications include close-proximity exploration of small bodies \citep{baresi2016bounded,lippe2020spacecraft} and formations along generic periodic orbits in the CR3BP \citep{gurfil2004stability}. Still, formation flying in close-proximity to tidally-locked satellites remains an understudied area.

For the case of Enceladus, the perturbation from Saturn and Enceladus' nonuniform gravity can produce strong differential accelerations on the spacecraft and destabilize the formation. Passively mitigating the differential effect of the perturbations during the design stage is crucial for formulating an efficient control plan that minimizes the overall cost of station-keeping. Active control may still be required to maintain the formation within operational requirements. Precise autonomous guidance, navigation, and control are required to ensure that the spacecraft can react to and counteract perturbations efficiently. Constant maneuvering can interfere with the continuity of the radio-tracking data. Thus, station-keeping maneuvers should be kept at a minimum when possible.

%  Fig.~\ref{Fig:std_range_rate} shows the standard deviation of the range-rate between the two spacecraft as a function of these two design parameters. The orbits of the spacecraft are propagated considering only degree~2 expansion of tidal potential including terms up to second order in eccentricity to isolate the differential effect of tides. The altitude of the second spacecraft is adjusted for each case to ensure that the formation remains stable and the spacecraft do not drift apart. Note that, for a given orbit altitude, the maximum angular separation is limited by the constraint on the link between the spacecraft not intersecting Enceladus. The results from this experiment indicate that the differential effect of tides causes an oscillatory response on the range rate and that the amplitude of the response for a given orbit altitude is maximized at specific angular separations. For altitudes below approximately 60~km the angular separation should be as large as visibility permits. For higher orbits, the optimal separation is around 70\textdegree. The global optimum in this case corresponds to an orbit altitude below 40~km, which may be limited by operational considerations.

% \begin{figure}[t]
%     \centering
%     \includegraphics{figures/std_range_rate.pdf}
%     \caption{Standard deviation in the inter-spacecraft range rate (in mm/s) between to spacecraft following circular equatorial orbits for different orbit altitudes and angular separations.\label{Fig:std_range_rate}}
% \end{figure}

\subsection{Mission simulations} % Ryan, so far this is pretty much what we had in the white paper.

Based on the aforementioned considerations, we conducted a series of mission simulations and performed detailed covariance analyses for two mission configurations: a single orbiter with radiometric tracking to the Earth and dual spacecraft (GRAIL-like configuration) with inter-satellite range-rate measurements. The covariance analysis is based on a least-squares principal and is a powerful tool for assessing the expected uncertainties in the estimated parameters \citep{park_detecting_2011,park_grail_2012,park_improved_2015}.

For simplicity, we assumed that both single and dual spacecraft are tracked continuously. In reality, the tracking of single spacecraft is limited by the Deep Space Network (DSN) availability. On interplanetary ranges, S-band (2.3 GHz), X-band (8.4 GHz) and Ka-band (32 GHz) have been used for spacecraft radio-tracking Doppler measurements. These bands respectively yield ranging accuracies of $\approx 5\cdot 10^{-7}$, $5\cdot 10^{-8}$ and $5\cdot 10^{-9}$ over a 10-s integration time \citep{bills2019simple,asmar2005spacecraft}. The impact of the frequency used for spacecraft Doppler tracking depends on the signal-to-noise ratio of the two-way radio link, which would depend on the spacecraft thermal noise, transmitter power, received signal noise, antenna type, etc. For the single spacecraft tracking scenario, the radio wave will go through the Earth media and Earth-to-Saturn distance encountering spatially and temporally variable solar plasma. On the other hand, for dual spacecraft tracking, the radio wave will only have to travel a few hundreds of km. In the latter case, the dual spacecraft tracking data quality would mainly be affected by the spacecraft thermal noise, which would be substantially better than in the single spacecraft case, i.e., increasing accuracy by at least a factor 10, which would yield at least an order of magnitude better result than the single spacecraft case.

We assumed an X-band tracking accuracy for the single spacecraft case, which is typical for a deep space mission. We only considered X-band for single spacecraft since Ka-band uplink capability is currently available only at the Goldstone DSN station. In addition, the Ka-band uplink capability is costly to maintain as it requires water vapor radiometer to get the full accuracy. If a Ka-band tracking of a single spacecraft is assumed, it can lead to a factor of 4 to 10 measurement accuracy accuracy improvement. 

\begin{figure}[ht]
\includegraphics[width=\columnwidth]{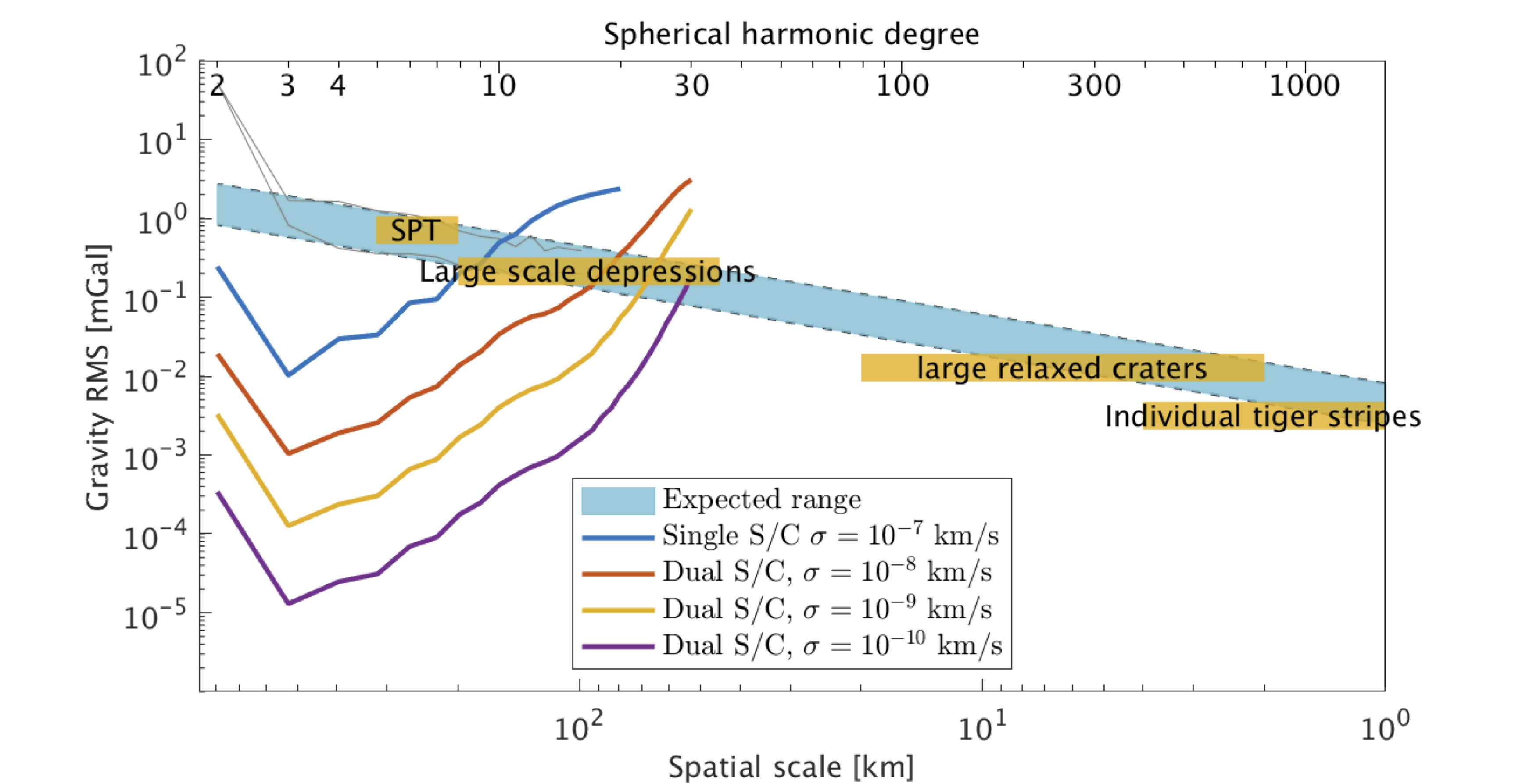}
\centering
\caption{Summary of the gravity anomaly signals due to various geologic landmarks. The vertical axis shows the Root-Mean-Square (RMS) magnitude of the gravity signal. The top horizontal axis shows spherical harmonic degree $n$. The bottom horizontal axis shows the corresponding spatial wavelength, $\lambda \approx 2\pi R/n$. The light blue region indicates the expected range of the gravity signal. The upper bound is given by a power law fitted to an uncompensated gravity-from-shape using the shape model by \cite{tajeddine_true_2017} with a shell density of 920~kg/m$^3$. The lower bound corresponds to gravity from compensated topography using the same shape model with a shell density of 600~kg/m$^3$. The large scale depressions refer to the chain of basins identified by \cite{tajeddine_true_2017}.}
\label{fig: ZuberPlot}
\end{figure}

We assumed a stable orbit around Enceladus with 60\textdegree{} inclination and periapsis and apoapsis altitudes of 150~km and 200~km, respectively. Both poles can be covered with flybys prior to being placed into this stable orbit, depending on the science requirements, which was not modeled in our simulations. We have simulated how accurately we can recover Enceladus' geophysical parameters, such as Love numbers, gravity field, and libration amplitude. The resulting radial gravity acceleration error spectra are shown in Fig. \ref{fig: ZuberPlot}. In addition, Fig. \ref{fig: ZuberPlot} shows the expected amplitude of the gravity signal as well as the spatial scales of several geologic features of interest. 

Assuming a single orbiter with two-way X-band tracking capability, and assuming the Doppler data accuracy of $10^{-7}$~km/s at a 60-s count time with a 28-day data collection duration, a gravity field up to degree $\approx$10 can be recovered and $k_2$ can be recovered with an accuracy of $\approx10^{-2}$. A high-resolution context imager with accurate pointing knowledge and image-motion-compensation capability, such as the Advanced Pointing Imaging Camera \citep{park_advanced_2020} (APIC), can measure $h_2$ and $l_2$ with accuracies $\approx$0.01 and $\approx$0.002, respectively. This corresponds to a relative uncertainty of the displacement Love numbers of 20-25\%. The libration amplitude can be recovered with an accuracy of $<$1~m at the equator (a relative uncertainty of $<$0.2\%). The accuracy of $h_2$, $l_2$, and libration amplitude can be improved with optimal distribution of crossover points \citep{park_improved_2015} or lower orbit altitude.

Recovering $h_2$ can further be achieved by laser altimetry \citep{steinbruegge2015galatides} or by radar in combination with stereo-imaging at crossover points \citep{steinbruegge2018reasontides}. Equivalent measurements are already planned with the Ganymede Laser Altimeter (GALA) \citep{hussmann2019gala} and the Radar for Europa: Ocean to Near-Surface (REASON) \citep{blankenship_reason_2018}. State of the art laser altimeters can achieve vertical resolutions of 10~cm \citep{araki2019performance}. However, the error budget for the measurement of radial tides is dominated by interpolation errors at cross-over points and orbit determination uncertainties of the spacecraft. Having an orbiter with an accurate radio science experiment is therefore beneficial for the $h_2$ measurement as well. Typical accuracies for such a configuration would be on the order of 1~m. The accuracy of the $h_2$ inversion then depends on the number of orbits $n_{orb}$ a spacecraft completes within its mission lifetime. Since the number of cross-over points increases with $n_{orb}^2$, millions of cross-over points can be collected over mission lifetimes of a few months. 

Using a radar sounder instead of a dedicated altimeter requires an additional stereo-camera to mitigate the ambiguity induced by surface clutter. The radar altimetric accuracy depends on the radar bandwidth, which is limited by half the center frequency. Ice is particularly transparent to frequencies between 1 and 300~MHz \citep{blankenship2009europabook}. To achieve a 3~m vertical resolution, a 50~MHz bandwidth is required, which is larger than the bandwidth used by typical radar sounders. However, by using stereo imaging in combination with radar sounders, resolutions up to a factor 4-5~better than the inherent range resolution can be achieved \citep{steinbruegge2018reasontides}. 

\begin{table}[htbp]
  \centering
    \begin{tabular}{cclllc}
    \toprule
          & \multicolumn{1}{p{6.25em}}{Range-rate accuracy [km/sec]} & \multicolumn{1}{c}{$k_{20}$} & \multicolumn{1}{c}{$k_{22}$} & \multicolumn{1}{c}{$k_{22}$} & \multicolumn{1}{p{5em}}{Global degree strength} \\
    \midrule
    \midrule
    single spacecraft & $10^{-7}$ & $9 \cdot 10^{-3}$ & $5 \cdot 10^{-1}$ & $4 \cdot 10^{-3}$ & 9 \\
    \midrule
    \multirow{3}[6]{*}{dual spacecraft} & $10^{-8}$ & $9 \cdot 10^{-4}$ & $2 \cdot 10^{-2}$ & $2 \cdot 10^{-4}$ & 17 \\
\cmidrule{2-6}          & $10^{-9}$ & $2 \cdot 10^{-4}$ & $3 \cdot 10^{-3}$ & $2 \cdot 10^{-5}$ & 23 \\
\cmidrule{2-6}          & $10^{-10}$ & $2 \cdot 10^{-5}$ & $4 \cdot 10^{-4}$ & $3 \cdot 10^{-6}$ & 30 \\
    \bottomrule
    \end{tabular}%
\caption{A summary of the potential Love numbers and gravity recovery. Standard deviations of the degree 2 Love numbers are shown. The degree strength refers to the spherical harmonic degree at which the gravity signal is equal to the noise in the gravity data. Thus, degree strength defines the global resolution of the gravity model.}
\label{tab: CovarianceAnalysisSummary}
\end{table}%

% \caption{A summary of the potential Love numbers recovery. Standard deviations of the degree-2 Love numbers are shown. The degree strength refers to the spherical harmonic degree at which the gravity signal is equal to the noise in the gravity data. Thus, degree strength defines the global resolution of the gravity model.}
% \label{tab: CovarianceAnalysisSummary}

A dual spacecraft architecture with inter-satellite Ka-band tracking (i.e., GRAIL-like scenario) was also considered for the same orbit configuration described above. We assumed range-rate accuracies of $10^{-8}$, $10^{-9}$, and $10^{-10}$~km/s. These or better (down to $3\cdot10^{-11}$~km/s) ranging accuracies were achieved by the GRAIL mission \citep{konopliv2013jpl}. Such measurement configuration can significantly improve the accuracy of the $k_2$ determination yielding $k_2$ error down to $\approx2 \cdot 10^{-5}$ (or relative accuracy of $\approx$0.1\%, Table \ref{tab: CovarianceAnalysisSummary}). The gravity field can be determined to degree~20 to 30 (see Fig. \ref{fig: ZuberPlot}), depending on the ranging accuracy. This level of accuracy would allow a definitive determination of the tidal phase lag and total tidal dissipation within Enceladus. The core viscoelastic moduli and ocean compressibility would remain virtually unconstrained for the studied mission configurations, although a lower bound on the core viscosity could be retrieved from the Love numbers if they are measured to an accuracy of 0.1\%.

Advances in small radios based on the IRIS transponder on the Mars CubeSat One mission and Artemis-1 CubeSats (e.g., software programmable Universal Space Transponder-Lite, or UST-Lite $\approx$1 kg, $\approx$15 W, two-way Allan deviation of $10^{-14}$ at 1000~s, \cite{pugh_universal_2017}) would enable a GRAIL-like scenario at little resource cost to the primary mission, but with less accurate satellite-to-satellite ranging accuracy. The CubeSat could be carried to the destination and network with the mothership via a deep space deployer providing power/thermal during cruise. This CubeSat would be enabled by new technologies in thermal management and propulsion subsystems. 

%% add ACS here

\section{Conclusions} % Anton

A focused geophysical investigation of Enceladus within a New Frontiers- or Flagship-class Enceladus Orbiter mission concept can address the Priority Science Questions outlined in this paper, as well as other compelling science questions. Thus, we conclude with the following recommendations:

{\bf{Recommendation 1:}} Geophysical measurements should be an essential component of future Enceladus exploration. Geophysical data can shed light on the mechanism of tidal dissipation and heat transport. Distinguishing between various locations of tidal dissipation can be achieved by mapping the variations of the total ice shell thickness, which can be further augmented by mapping the elastic shell thickness either by localizing gravity-topography admittance or by measuring flexural profiles at various locations. Thermal IR mapping away from the South Polar Terrain is needed to estimate the global heat flux. Laser or radar altimetry could be used to derive a high-resolution shape model and measure tidal deformations.

{\bf{Recommendation 2:}} A dedicated gravity mapping investigation should be considered for Enceladus. A GRAIL-like mission would be able to measure tidal phase lag and significantly improve the accuracy of the gravity-topography admittance. This would enable investigating the energetics of Enceladus, thus establishing a foundation for understanding Enceladus' long-term habitability. That investigation could be added to New-Frontiers or Flagship-class mission concepts to Enceladus for little additional resource requirements by leveraging current and upcoming developments in CubeSat technologies and small radios.

\section*{Acknowledgements}
A portion of this research was carried out at the Jet Propulsion Laboratory, California Institute of Technology, under a contract with the National Aeronautics and Space Administration (80NM0018D0004). 
\section*{Appendix}
\renewcommand{\thesubsection}{\Alph{subsection}}

\subsection{Markov chain Monte-Carlo internal structure inversion}\label{sec: MCMC}

We use the affine-invariant ensemble sampler \citep{goodman_ensemble_2010} implemented in the publicly available emcee python library \citep{foreman-mackey_emcee_2013}. First, an ensemble of internal structure models was generated by sampling the prior probability distribution of the model parameters. An ensemble consists of individual model realizations or so-called walkers. Typically, from 500 to 1500 walkers were used. More walkers were used for the MCMC runs with more stringent measurement errors. Affine-invariant ensemble sampler uses walkers from the previous step in the Markov chain to generate the positions at the following step. The walker positions are updated based on the likelihood function. The likelihood function tells how well a model reproduces the observations. We use the likelihood function in the following form:

\begin{equation}
\log L \propto -\frac{1}{2}(\bf{X} - {\bf{Y}})^T \Sigma^{-1} (\bf{X} - \bf{Y})
\label{eq: LogL}
\end{equation}

\noindent where $\bf{X}$ is the vector of observations and $\bf{Y}$ is the vector of model predictions, which includes degree~2 gravity and shape coefficients, Love numbers and libration amplitude at the orbital frequency. $\bf{\Sigma}$ is the covariance matrix that contains contributions from the observational and model covariances: $\bf{\Sigma}={{\bf{\Sigma}}}_{\rm{model}}+{\bf{\Sigma}}_{\rm{obs}}$. The observation covariance is given by the gravity, shape, tidal Love numbers and libration determination accuracies. The observations were assumed independent, thus yielding a diagonal ${\bf{\Sigma}}_{\rm{obs}}$. Each Markov chain was inspected visually and the initial burn-in steps were discarded. The Markov chains were run until convergence as informed by visual inspection and computed chain auto-correlation times.

 % ${\bf{X}} = \{C_{20}, C_{22}, S_{22}, k_{2}, h_{2}, l_{2}, Z_2, Z_3, ..., Z_n, \gamma, \}$
% {\bf{\Sigma}}_{model} model describe 
% Autocorrelation times were computed using
% Each Markov chain was inspected visually and the initial burn in steps were discarded
% 
\bibliography{Bibliography_Enceladus}
\bibliographystyle{aasjournal}



\end{document}